%% file: main.tex
\documentclass[acmsmall,screen]{acmart}
\usepackage{xspace}
\usepackage{xcolor}
\usepackage{enumitem}
\usepackage{pgfplots}
\pgfplotsset{compat=1.18} % 可根据你使用的版本调整
\usetikzlibrary{patterns}
\usepgfplotslibrary{statistics}
\usepackage{multirow}
\usepackage{subfig}
\usepackage{pifont}
\usepackage{tikz}
\usepackage[framemethod=TikZ]{mdframed}
\usepackage[most]{tcolorbox}
\usepackage{fontawesome5}

\definecolor{bg}{HTML}{F1F7EC}
\definecolor{bgline}{HTML}{D6E6C7}

\newmdenv[
    roundcorner=2pt,        % 圆角半径
    linewidth=1pt,          % 边框粗细
    linecolor=bgline,% 边框颜色
    backgroundcolor=bg, % 背景色
    leftmargin=0pt,         % 左外边距
    rightmargin=0pt,        % 右外边距
    innerleftmargin=6pt,    % 左内边距
    innerrightmargin=6pt,   % 右内边距
    innertopmargin=4pt,     % 上内边距
    innerbottommargin=4pt,  % 下内边距
    skipabove=\topsep,      % 上间距
    skipbelow=\topsep,      % 下间距
    needspace=3\baselineskip % 分页保护
]{finding}

\newcommand{\findingtitle}[1]{\textbf{Finding #1.}\xspace}

\AtBeginDocument{%
  }

\setcopyright{acmlicensed}
\copyrightyear{2026}
\acmYear{2026}
\acmDOI{XXXXXXX.XXXXXXX}

\acmISBN{978-1-4503-XXXX-X/2018/06}

\newcommand{\mybench}{\textsc{DesBench}\xspace}
\newcommand{\passk}{\textit{Pass@k}\xspace}
\newcommand{\compk}{\textit{Compilation@k}\xspace}
\newcommand{\passo}{\textit{TPass\%}\xspace}
\newcommand{\compo}{\textit{TCompilation\%}\xspace}
\newcommand{\preci}{\textit{Prec}\xspace}
\newcommand{\recal}{\textit{Recall}\xspace}
\newcommand{\fone}{\textit{F1}\xspace}
\newcommand{\clscov}{CCov}
\newcommand{\methcov}{MCov}
\newcommand{\linecov}{LCov}
\newcommand{\classmatch}{$Match_{class}$\xspace}
\newcommand{\methodmatch}{$Match_{method}$\xspace}
\setlist[itemize]{leftmargin=*}

\begin{document}

\title{\textit{From What to How}: Bridging User Requirements with Software Development Using Large Language Models
}

%%
%% The "author" command and its associated commands are used to define%% the authors and their affiliations.%% Of note is the shared affiliation of the first two authors, and the%% "authornote" and "authornotemark" commands

\author{Xiao He}
\email{hexiao@ustb.edu.cn}%
\affiliation{% School of Computer and Communication Engineering, 
  \institution{University of Science and Technology Beijing}
  \city{Beijing}
  \country{Beijing, China}
}

\author{Ru Chen}
\email{chenru@ustb.edu.cn}
\affiliation{% School of Computer and Communication Engineering,
  \institution{University of Science and Technology Beijing}
  \city{Beijing}
  \country{Beijing, China}
}
\author{Jialun Cao}
\email{jialuncao@ust.hk}
\authornote{Corresponding author.}
\affiliation{%
  \institution{Hong Kong University of Science and Technology}
  \city{Hong Kong}
  \country{Hong Kong, China}
}

\renewcommand{\shortauthors}{\mybench}

\input{sections/abstract}

\begin{CCSXML}
<ccs2012>
   <concept>
       <concept_id>10011007.10011074.10011075.10011077</concept_id>
       <concept_desc>Software and its engineering~Software design engineering</concept_desc>
       <concept_significance>500</concept_significance>
       </concept>
   <concept>
       <concept_id>10011007.10011074.10011075.10011079.10011080</concept_id>
       <concept_desc>Software and its engineering~Software design techniques</concept_desc>
       <concept_significance>500</concept_significance>
       </concept>
   <concept>
       <concept_id>10011007.10011074.10011092.10011782</concept_id>
       <concept_desc>Software and its engineering~Automatic programming</concept_desc>
       <concept_significance>500</concept_significance>
       </concept>
 </ccs2012>
\end{CCSXML}

\ccsdesc[500]{Software and its engineering~Software design engineering}
\ccsdesc[500]{Software and its engineering~Software design techniques}
\ccsdesc[500]{Software and its engineering~Automatic programming}

\keywords{Large Language Model, Software Design, Code Generation, Test Case Generation}

\received{20 February 2007}
\received[revised]{12 March 2009}
\received[accepted]{5 June 2009}

\maketitle

\input{sections/introduction-new}
\input{sections/benchmark}

\input{sections/studydesign}
\input{sections/results}
\input{sections/relatedwork}
\input{sections/conclusion}

% \begin{acks}
% \end{acks}

\bibliographystyle{ACM-Reference-Format}
\bibliography{main}


\end{document}

%% file: sections/abstract.tex
\begin{abstract}

Recently, large language models (LLMs) are extensively utilized to enhance development efficiency, leading to numerous benchmarks for evaluating their performance. 
However, these benchmarks predominantly focus on implementation, overlooking the equally critical aspect of software design. 
This gap raises two pivotal questions: \textbf{
(1) Can LLMs handle software design? (2) Can LLMs write code following the specific designs?
}
To investigate these questions, this paper proposes \mybench, a design-aware benchmark for evaluating LLMs on three software design-related tasks: design-aware code generation, object-oriented modeling, and the design of acceptance test cases. 
\mybench comprises 30 \textit{manually} crafted Java projects that include requirement documents, design models, implementations, and acceptance tests, amounting to a total of 30 design models, 194 Java classes, and 737 test cases.
We evaluated seven state-of-the-art LLMs, including three DeepSeek R1, two Qwen2.5, and two GPT models, using \mybench. 
The results reveal that LLMs remain significantly challenged by the intricacies of software design:
(1) For code generation, LLMs struggle to produce correct implementations when provided with only high-level or no designs.
(2) In object-oriented modeling, while LLMs can accurately identify objects and classes, they face challenges in defining operations and inter-class relationships.
(3) Acceptance test cases generated by LLMs from functional requirements achieve code coverage quality comparable to those written by humans.
Our research highlights the current limitations of LLMs in managing software design and calls for further investigation into new design methodologies and languages suitable for LLM-based development.

\end{abstract}

%% file: sections/introduction-new.tex
\section{Introduction}

The proliferation of Large Language Models (LLMs) has significantly enhanced development efficiency across various industries. Research from entities like McKinsey reveals that one-third of enterprises now utilize generative AI technologies, with 40\% planning to escalate their AI investments~\cite{role-ai2023}. 
As recent research~\cite{role-ai-25} on AI applications highlights, to maximize the usefulness of AI in Requirement Engineering (RE)~\cite{role-ai-25,formal-re2025,cheng2024generative} processes, it is necessary to systematically identify the most suitable ways to use it in the software development lifecycle.

Imagine a scenario where future users leverage LLMs to seamlessly navigate the software development lifecycle, from articulating needs to implementation. For instance (Figure~\ref{fig:DesBench}), users might express a requirement such as ``\textit{I need an airline management system}''. These users ask LLMs to help design, develop, and test the implemented software. This scenario highlights a critical need: simplifying complex software development processes for \textbf{\textit{non-expert users}}. They know what they want (the \textbf{\textit{what}}), but often lack the technical expertise to determine how to achieve it (the \textbf{\textit{how}}).

Existing benchmarks remain predominantly \textit{implementation-centric}, emphasizing code generation against  specifications for implementation---the transformation from \textbf{\textit{how}} to code, instead of 
from \textbf{\textit{what}} to code. Figure \ref{fig:devbench_projecteval} presents two example requirements from~\cite{DevEval-BowenLi-etal-2025-prompting, projecteval}. Clear that these requirements already provide technical (e.g., \texttt{IdWorker.java}, \texttt{SidWorker.java}) and implementation (e.g., 
decomposing requirements into checklists and code skeletons) details, going beyond the \textit{what}.

Therefore, this paper fills this gap by introducing \mybench, which consists of 30 cases that start from a user requirement articulating \textit{what}, and progress through \textit{how} to design, implement, and test a software system, ensuring a seamless transition from \textit{what} to \textit{how}. 
Figure~\ref{fig:DesBench} illustrates a streamlined case in \mybench, which closely mirrors real-world application scenarios. To begin with, the user \faUser~requests the development of an airline management system tasked with managing flight scheduling and tracking. In response, software designers \faUserCog~ design the system in two levels: {high-level design} outlines the system's architecture using class diagrams that specify classes, attributes, operations, and relationships, including inheritance and associations/aggregations; {low-level design} details the implementation, usually including code skeletons that define fields and function interfaces for each class. Developers \faUserEdit~ then code according to these designs, while testers \faUserLock~ ensure the system meets user requirements through acceptance testing.

\input{figureTex/case}
\input{figureTex/requirement-case}

\mybench consists of 30 Java projects that span 24 business domains, containing 194 classes and 737 test cases (see details in Table~\ref{fig:general-info}).
Each project is complemented with a {domain description} detailing the background of the domain, 3--6 {functional requirements}, a {test specification} outlining 15--31 test cases in natural language (NL), along with a {domain model} and a {design model}, both represented as class diagrams. Moreover, each project provides a {reference implementation} and JUnit \textit{test cases} that implement the test specification.
We carefully curated each case, ensuring the quality of both the requirements in natural language and the code in Java. Finally, since we curated all requirements and code, the risk of data contamination was minimized.

\input{table/benchmark-details}

For evaluation, we assessed seven popular LLMs against \mybench. 
The experiments have yielded interesting findings. First,  LLMs usually can successfully identify objects from requirements with high precision and thus create classes correctly (with F1 up to 0.85), while struggling with reasoning about the correct operations and relations among objects. 
Second, LLMs prefer low-level designs (e.g., code skeletons) to high-level designs (e.g., class diagrams). 
Third, LLMs generate acceptance test cases from functional requirements that achieve code coverage quality comparable to those written by humans.
Yet, they tend to generate too many test cases within one test method.

The contributions of this paper are summarized below.
\begin{itemize}
    \item \textbf{Novelty.} Introducing \mybench, the first benchmark designed to evaluate LLMs' ability to derive OO designs directly from user requirements and crucially adhere to these designs during code generation---an aspect overlooked by existing implementation-centric benchmarks---by utilizing implementation-agnostic NL requirements that accurately simulate real-world scenarios where requirements specify the \textit{what} without prescribing the \textit{how}.

    \item \textbf{Significance.} \mybench enables granular evaluation by combining varying design abstraction levels. For example, for code generation, it spans from NL requirements to those enhanced with method signatures, class diagrams, or code skeletons. 
    This systematically measures how incremental design detail affects LLMs’ coding and testing performance.

    \item \textbf{New insight.} Our analysis reveals a major limitation in LLM software design abilities: they have difficulty crafting high-level designs and converting these into correct code. 
\end{itemize}

%% file: figureTex/case.tex
\begin{figure*}[!tb]
    \centering
    \includegraphics[width=1\linewidth]{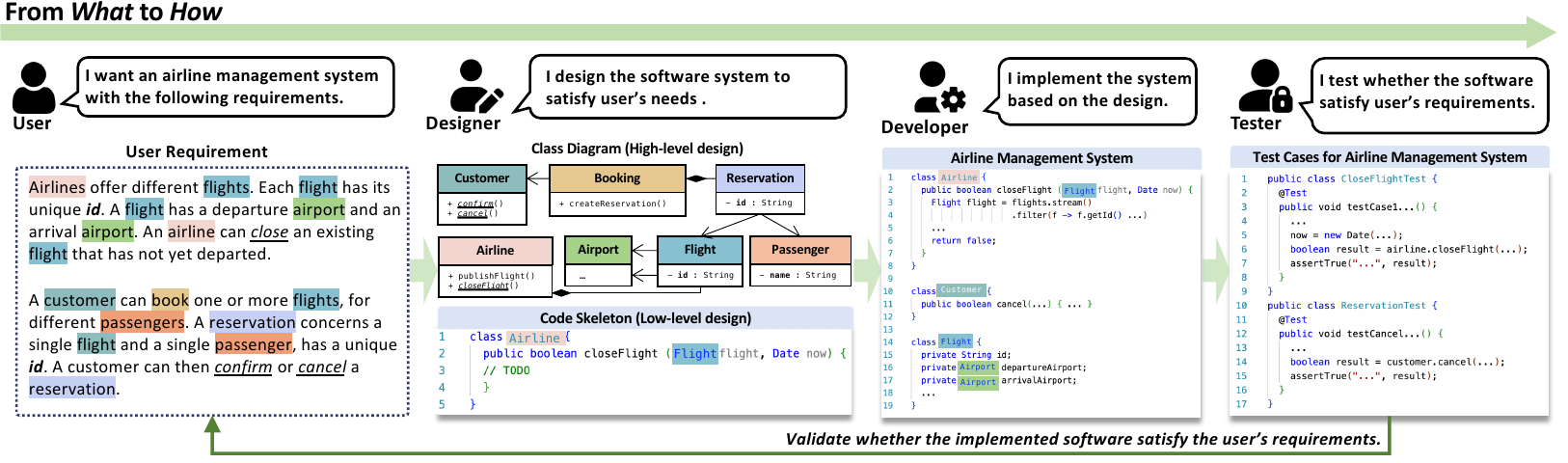}
    \caption{The software lifecycle: from requirements to code}
    \label{fig:DesBench}
\end{figure*}

%% file: figureTex/requirement-case.tex
\begin{figure*}[!thb]
    \centering
    \includegraphics[width=1\linewidth]{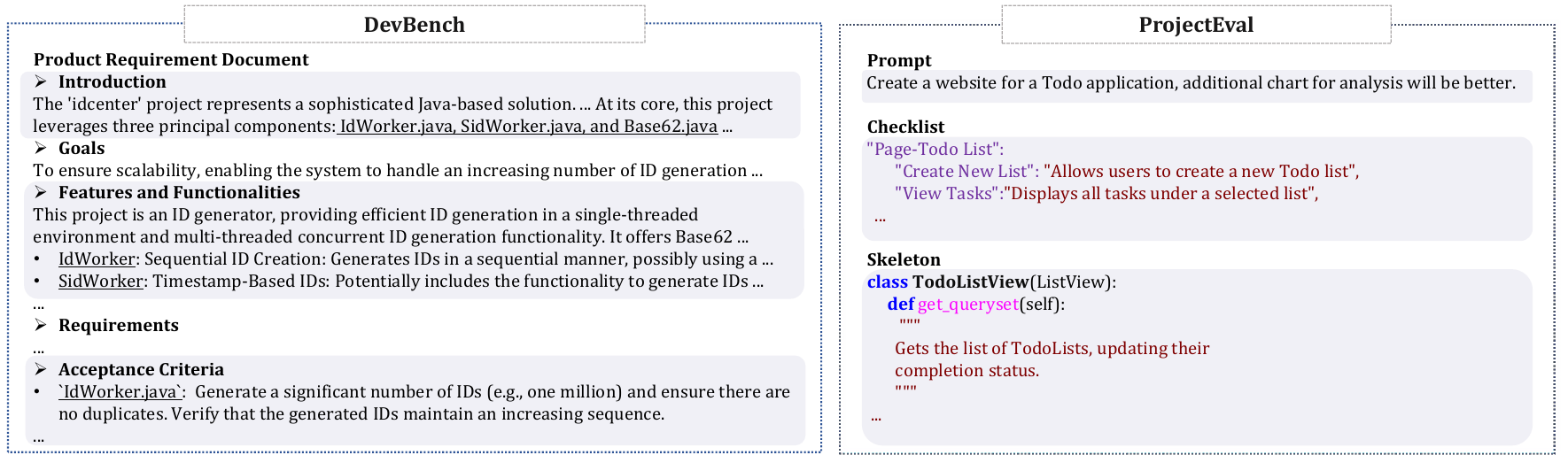}
    \caption{Examples Requirements from DevBench \cite{deveval} and ProjectEval \cite{projecteval}}
    \label{fig:devbench_projecteval}
\end{figure*}

%% file: table/benchmark-details.tex
\begin{table}[!htb]
\centering
\caption{Project Statistics in \mybench}\label{fig:general-info}
\renewcommand\arraystretch{1.25}
{\scriptsize{
% \usepackage{multirow}
% \usepackage{multirow}
\resizebox{1\linewidth}{!}{
\begin{tabular}{l|l|lrr|rrrr|rrr|rrr}
\toprule
\multicolumn{1}{l|}{\multirow{2}{*}{\textbf{Name}}} & \multicolumn{1}{l|}{\multirow{2}{*}{\textbf{Domain}}} & \multicolumn{1}{l}{\multirow{2}{*}{\textbf{Source}}} & \multicolumn{1}{r}{\multirow{2}{*}{\textbf{WoR}}} & \multicolumn{1}{r|}{\multirow{2}{*}{\#\textbf{FR}}}  & \multicolumn{4}{c|}{\textbf{Design} Model}                                                                               & \multirow{2}{*}{\textbf{LoC}} & \multirow{2}{*}{\#\textbf{TC}} & \multirow{2}{*}{\textbf{LoT}} & \multicolumn{3}{c}{\textbf{Coverage}}                                                      \\
\multicolumn{1}{l|}{}                      & \multicolumn{1}{l|}{}                        & \multicolumn{1}{c}{}                & \multicolumn{1}{l}{}        &                         & \multicolumn{1}{l}{\#Cls} & \multicolumn{1}{l}{\#Op} & \multicolumn{1}{l}{\#Inh} & \multicolumn{1}{l|}{\#As} &                      &                          &                      & \multicolumn{1}{l}{Cls} & \multicolumn{1}{l}{Meth} & \multicolumn{1}{l}{Line} \\
\midrule
D01\_CIN              & \multirow{2}{*}{Cinema}                  & \cite{nlpset}                  & 447                                        & 5                      & 5                         & 7                        & 0                         & 5                        & 209                  & 25                       & 395                  & 1.00                      & 0.67                       & 0.70                     \\
D02\_CINS             &                   & \cite{nlpset}                  & 287                                        & 6                      & 4                         & 6                        & 0                         & 4                        & 135                  & 31                       & 505                  & 1.00                      & 0.77                       & 0.80                    \\ 
\hline

D03\_AP               & \multirow{2}{*}{Course}                  & \cite{nlpset}                  & 254                                        & 5                      & 6                         & 11                       & 0                         & 5                        & 312                  & 25                       & 484                  & 1.00                      & 0.48                       & 0.58                     \\
D04\_UES              &                    & \cite{mlset}                  & 258                                        & 5                      & 5                         & 5                        & 0                         & 3                        & 148                  & 25                       & 335                  & 1.00                      & 0.52                       & 0.72                     \\

\hline

D05\_FM               & \multirow{2}{*}{Document}                & \cite{nlpset}                  & 174                                        & 5                      & 6                         & 8                        & 3                         & 3                        & 142                  & 25                       & 363                  & 1.00                      & 0.73                       & 0.85                     \\
D06\_DS               &                 & \cite{nlpset}                  & 265                                        & 5                      & 7                         & 9                        & 4                         & 4                        & 112                  & 25                       & 382                  & 1.00                      & 0.92                       & 0.87                     \\
\hline

D07\_CWH              & \multirow{2}{*}{House}                   & \cite{nlpset}                  & 296                                        & 5                      & 5                         & 5                        & 0                         & 4                        & 137                  & 25                       & 535                  & 1.00                      & 0.92                       & 0.94                     \\
D08\_TBH              &                   & \cite{nlpset}                  & 326                                        & 5                      & 9                         & 8                        & 5                         & 6                        & 235                  & 25                       & 283                  & 1.00                      & 0.73                       & 0.71                     \\
D09\_LS               & \multirow{3}{*}{Library}                  & \cite{mlset}                  & 121                                        & 4                      & 5                         & 4                        & 0                         & 4                        & 189                  & 20                       & 504                  & 1.00                      & 0.66                       & 0.75                     \\

\hline

D10\_OLRS             &                   & \cite{mlset}                  & 347                                        & 5                      & 10                        & 5                        & 2                         & 2                        & 170                  & 25                       & 428                  & 1.00                      & 0.97                       & 0.95                     \\
D11\_MLIB             &                  & \cite{nlpset}                  & 732                                        & 5                      & 9                         & 8                        & 5                         & 4                        & 195                  & 25                       & 425                  & 1.00                      & 0.74                       & 0.83                     \\

\hline 

D12\_ALF              & Airline              & \cite{nlpset}                  & 588                                        & 6                      & 10                        & 12                       & 0                         & 10                       & 358                  & 30                       & 666                  & 1.00                      & 0.70                       & 0.80                     \\
D13\_PETS             & Animal                  & \cite{nlpset}                  & 192                                        & 5                      & 4                         & 6                        & 2                         & 1                        & 84                   & 25                       & 297                  & 1.00                      & 0.76                       & 0.87                     \\
D14\_OAGS             & Art auction              & \cite{mlset}                  & 167                                        & 3                      & 6                         & 3                        & 0                         & 2                        & 194                  & 15                       & 261                  & 1.00                      & 0.69                       & 0.74                     \\
D15\_BKS              & Bank                    & \cite{nlpset}                  & 420                                        & 5                      & 5                         & 9                        & 2                         & 2                        & 172                  & 25                       & 351                  & 1.00                      & 0.64                       & 0.77                     \\
D16\_RCGS             & Car rental              & \cite{nlpset}                  & 276                                        & 5                      & 5                         & 6                        & 0                         & 6                        & 190                  & 25                       & 521                  & 1.00                      & 0.73                       & 0.83                     \\
D17\_OPMS             & Company                  & \cite{mlset}                  & 204                                        & 4                      & 11                        & 6                        & 4                         & 6                        & 208                  & 20                       & 489                  & 1.00                      & 0.70                       & 0.78                     \\
D18\_DEPT             & Department              & \cite{nlpset}                  & 185                                        & 5                      & 4                         & 9                        & 0                         & 5                        & 129                  & 25                       & 339                  & 1.00                      & 1.00                       & 0.96                     \\
D19\_EMS              & Employee                & \cite{nlpset}                  & 297                                        & 5                      & 9                         & 5                        & 5                         & 5                        & 280                  & 25                       & 346                  & 1.00                      & 0.54                       & 0.67                     \\
D20\_BWA              & Exercise                & \cite{nlpset}                  & 448                                        & 5                      & 7                         & 7                        & 0                         & 6                        & 208                  & 25                       & 408                  & 1.00                      & 0.75                       & 0.85                     \\
D21\_OFDS             & Food                     & \cite{mlset}                  & 163                                        & 4                      & 6                         & 4                        & 2                         & 3                        & 150                  & 20                       & 365                  & 1.00                      & 0.70                       & 0.73                     \\
D22\_FT               & Football                & \cite{nlpset}                  & 283                                        & 5                      & 5                         & 5                        & 0                         & 5                        & 172                  & 25                       & 588                  & 1.00                      & 0.74                       & 0.86                     \\
D23\_IPOA             & IPO                     & \cite{nlpset}                  & 421                                        & 5                      & 6                         & 9                        & 0                         & 5                        & 228                  & 26                       & 540                  & 1.00                      & 0.79                       & 0.84                     \\
D24\_IS               & Information             & \cite{nlpset}                  & 274                                        & 5                      & 4                         & 7                        & 0                         & 3                        & 144                  & 25                       & 329                  & 1.00                      & 0.85                       & 0.87                     \\
D25\_MAIL             & Mail delivery           & \cite{nlpset}                  & 260                                        & 5                      & 6                         & 6                        & 2                         & 4                        & 145                  & 25                       & 571                  & 1.00                      & 0.84                       & 0.83                     \\
D26\_MS               & Music rental            & \cite{nlpset}                  & 297                                        & 5                      & 6                         & 13                       & 0                         & 10                       & 222                  & 25                       & 412                  & 1.00                      & 0.69                       & 0.72                     \\
D27\_OPRS             & Paper review             & \cite{mlset}                  & 260                                        & 5                      & 9                         & 10                       & 3                         & 4                        & 208                  & 25                       & 399                  & 1.00                      & 0.63                       & 0.73                     \\
D28\_SCH              & School                  & \cite{nlpset}                  & 323                                        & 5                      & 8                         & 11                       & 2                         & 7                        & 167                  & 25                       & 336                  & 1.00                      & 0.74                       & 0.80                     \\
D29\_ORS              & Trip                     & \cite{mlset}                  & 453                                        & 5                      & 8                         & 12                       & 2                         & 5                        & 345                  & 25                       & 504                  & 1.00                      & 0.63                       & 0.69                     \\
D30\_VR               & Video rental            & \cite{nlpset}                  & 451                                        & 5                      & 4                         & 6                        & 0                         & 4                        & 183                  & 25                       & 384                  & 1.00                      & 0.74                       & 0.84                     \\
\hline
\textbf{Total}      & 24 domains &       &  9469 &   147 &   194     & 222 & 43 & 137 & 5771 & 737 & 12745 & 1.00 & 0.73 & 0.80\\
\bottomrule
\multicolumn{15}{l}{\textbf{FR}: functional requirements; \textbf{WoR}: words of requirement; \textbf{LoC}: lines of code; \textbf{LoT}: lines of tests; \textbf{Cls}: classes; \textbf{Op}: core operations;} \\
\multicolumn{15}{l}{\textbf{Inh}: inheritances; \textbf{As}: associations/aggregations; \textbf{TC}: test cases; \textbf{Meth}: method}\\
\end{tabular}
}
}}
\end{table}

%% file: sections/benchmark.tex
\section{Benchmark Construction}
% Format
\subsection{Benchmark Format}
\begin{figure*}[!thb]
    \centering
    \includegraphics[width=1\linewidth]{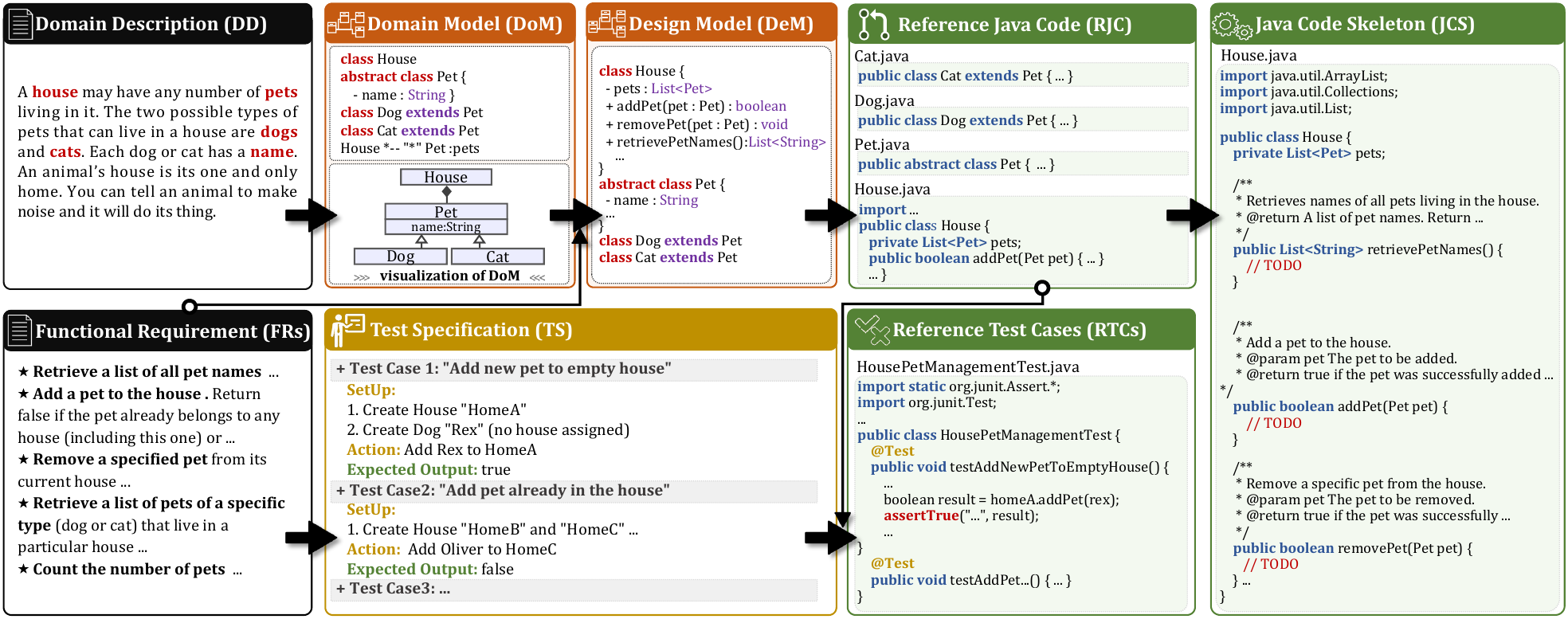}
    \caption{Structure of a single project in \mybench}
    \label{fig:structure}
\end{figure*}

\mybench comprises 30 projects.
Each project, as shown in Figure \ref{fig:structure}, includes the following artifacts.
\begin{itemize}
    \item The \textit{domain description} (DD) provides an overview of the application domain by defining entities and their relationships, intentionally excluding any design or implementation specifics. %, e.g., the classes, fields, and methods that must be developed.

    \item The \textit{functional requirements} (FRs) specify the business logic to be implemented. 

    \item The \textit{test specification} (TS) details a set of acceptance test cases composed in natural language. 

    \item The \textit{domain model} (DoM) is a class diagram encoded in PlantUML.
    DoM is derived primarily from DD, emphasizing a comprehensive understanding of the domain. 

    \item The \textit{design model} (DeM) is a refined PlantUML class diagram derived from the DoM, taking into account FRs to offer further programming guidance. 

    \item The \textit{reference Java code} (RJC) is the implementation of DD and FR, also consistent with DeM.

    \item  The \textit{reference test cases} (RTCs) realize TS for RJC under the JUnit framework.

    \item The \textit{Java code skeleton} (JCS) is created by removing the bodies of core methods in RJC.
\end{itemize}

% Construction Process
\subsection{Construction Procedure}
The construction process of \mybench is explained as follows.

\noindent \textbf{Step 1}.
Initially, we gathered 183 descriptions from the literature \cite{chen23,nlpset,mlset} within the domains of model-driven engineering and requirement engineering. These descriptions were initially used to evaluate methodologies pertinent to object-oriented domain modeling and were not tied to any specific implementation. We then eliminated descriptions based on the following criteria: (1) duplicates, (2) excerpts that were incomplete or taken from longer documents, (3) content that was not applicable to domain modeling, such as those focusing on communication protocols or user interfaces, (4) descriptions that did not align with object-oriented principles, and (5) descriptions that were either too simplistic or overly complex. Ultimately, we chose 30 domain descriptions (DDs) and meticulously revised them to correct typographical and formatting errors, unify terminology, enhance clarity to reduce interpretive ambiguities, and remove extraneous information.

\noindent \textbf{Step 2}.
For each DD, we crafted 3--6 functional requirements (FRs) that encompass object CRUD operations. To ensure non-trivial FRs, each must link at least two domain entities, requiring navigation of various objects and access to multiple attributes. To avoid confusion, we explicitly defined formats for complex values, e.g., a \textit{Date} should use the ``yyyy-MM-dd'' format. We also clearly specify constraints in our requirements, indicating if boundary values are inclusive, as terms like ``within'' or ``between'' can lead to misunderstandings.
We created 5--6 NL test cases per FR using boundary value analysis, resulting in a testing specification (TS) for each project. These test cases validate FRs, detailing test inputs (format and value), setups, expected outputs, and assertion logic. Implementation hints, like class/method names, are excluded in TSs, favoring DD vocabulary.

\noindent \textbf{Step 3}.
We applied OO modeling to DDs, yielding 30 domain models (DoMs) that define domain classes and their relationships.
Further, we derived 30 design models (DeMs).
For each association in a DoM, one or two class attributes were introduced into the DeM to depict association implementation, based on whether it is unidirectional or bidirectional.
\textit{Core operations} were defined within DeMs by decomposing FRs into high-cohesion operations following design principles. 
These OOD models were reviewed by two additional authors to verify their consistency with DDs.

\noindent \textbf{Step 4}.
We handcrafted Java classes as the reference implementations (RJCs) to satisfy DDs and FRs based on DeMs. 
We guarantee that every domain entity and FR is realized. 
These Java classes refine DeMs by potentially introducing new fields and methods, such as getter/setter methods.

\noindent \textbf{Step 5}.
For RJC, we developed JUnit test cases (RTCs) to realize TSs.
We tested RJC with RTCs and achieved an 100\% class coverage, 73\% method coverage, and 80\% line coverage.

\noindent \textbf{Step 6}.
We created Java code skeletons by replacing the \textit{core method} bodies in RJC with a ``\verb|// TODO|'' comment. The core methods correspond to core operations in DeM linked to FRs, whereas simple methods like \textit{getters} and \textit{setters} remain intact. 
\mybench retains broad descriptions for classes and methods but omits detailed algorithmic outlines. 
Our goal is to assess the design skills of LLMs, focusing on their ability to autonomously devise code structures, define responsibilities, establish boundaries, and recognize inter-dependencies among classes and methods, through DDs, FRs, and DeMs, without specific coding-level instructions.

%% file: sections/studydesign.tex
\section{Experiment Design}\label{sec:design}

\subsection{Research Questions}
With \mybench, we perform an experiment to assess current LLMs in terms of their design abilities across different development scenarios by answering the following research questions:
\begin{itemize}
    \item \textbf{RQ1 (Design-aware code generation):} How effectively do LLMs perform design-aware code generation?
    We provide LLMs with DDs and FRs, DeMs, asking them to generate Java code. 
    This evaluates the LLMs' basic ability to understand and utilize software design for code generation.

    \item \textbf{RQ2 (Object-Oriented Design):} How adept are LLMs at creating class diagrams from requirements?
    We explore their ability to perform object-oriented design based on DDs and FRs.

    \item \textbf{RQ3 (Acceptance test generation):} To what extent can LLMs design and generate acceptance test cases to validate the implementation against FRs? 
\end{itemize}

\subsection{Metrics}
For RQ1, we adopted two probabilistic metrics, i.e., \passk\cite{HumanEval-passk} and \compk \cite{JavaBench10.1145/3691620.3695470}, to measure the generated code. 
\passk denotes the probability of at least one of the top-k generated solutions passing all tests, while \compk signifies the probability of at least one of the top-k generated solutions compiling successfully.
For each project, $n$ is the number of solutions that are sampled from an LLM, and $c$ denotes the number of correct solutions. 
We set $n$ to 10 and $k$ to 1, 3, and 5.

Moreover, to assess the alignment between LLMs' code design and our design, we define two metrics, \classmatch and \methodmatch, calculated as in Equation \ref{eq:codematch}, where the term \textit{item} refers to classes or methods. For example, if there are 2 classes in RJC, but only 1 class name in the generated code matches those in RJC, then \classmatch is 0.5.
{\small{
\begin{equation}
    Match=\underset{projects}{\mathbb{E}}\left[ \frac{\text{\# of generated items that match the item names in RJC}}{\text{\# of items in RJC}} \right]    \label{eq:codematch}
\end{equation}
}}%

For RQ2, we used the metrics of \preci, \recal, and \fone to assess generated OO models.
An OO model contains the following types of elements, including classes, attributes, operations, inheritances, and associations/compositions.
By matching generated models with Oracle models provided in \mybench, we calculate these metrics for every element type.

In RQ3, we use two code metrics, i.e., \compo and \passo, and three coverage metrics, i.e., \clscov, \methcov, and \linecov, to assess the quality of generated test cases.
\compo is the likelihood that the generated test cases are compilable, while \passo means the likelihood that the reference Java code can pass the generated code. 
Ensuring that the correct code successfully passes the generated test cases is a \textit{necessary condition} for effective test case generation. A lower \passo indicates poorer quality in the generated tests.
The metrics \clscov, \methcov, and \linecov, denoting class/method/line coverage are used to assess the adequacy of generated tests.

\subsection{Experiment Setup}
% This section elaborates on the experimental processes and settings for each RQ.

For {\textbf{RQ1}} which assesses LLM's capability in comprehending, deriving, and employing software designs during code generation, we provide them with DDs, FRs, and additional design information, and request that they produce source code that meets the requirements.
For each project, we ask LLMs to return 10 answers.
Then, we compile the generated code and use RTCs for testing.
Finally, we calculate \compk and \passk.
We also match class and method names in the generated code with those defined in RJC, and calculate \classmatch and \methodmatch.
Table \ref{tab:rq1-configs} shows the concrete input settings of RQ1.
Each setting evaluates how design information 
% (from raw to high/low level)
influences code generation.
Starting with a baseline setting (`DD+FR'), we progressively add method names, high-level (class diagrams), low-level (skeletons), and combined designs to evaluate how design richness guides code structure and quality. 
Comparing these settings via \compk and \passk identifies optimal design inputs for improving LLM performance.

\input{table/RQ1-setup}

For {\textbf{RQ2}}, we request LLMs to generate one class diagram (in PlantUML format) from DD and FRs for each project to assess LLMs' capability of performing OO design.
After generation, two authors collaboratively compare generated models with DoMs/DeMs, calculating \preci, \recal, and \fone for classes, attributes, operations, inheritances, and associations/compositions.
We pair model elements based on their names, considering word variations and synonyms.
When matching operations, we ignore trivial getter and setter operations and focus on core operations that correspond to the FRs.
The design task is performed based on three settings as shown in Table \ref{tab:rq1-configs}.

For \textbf{RQ3}, LLMs are required to generate acceptance test cases for reference Java code according to FRs using the following three settings:
\begin{itemize}
    \item \textbf{FR+RJC}---Generate test cases based on FRs and RJC. 
    LLMs are permitted to design test cases freely. We ask that the number of test cases generated by LLMs matches the number of RTCs.
    
    \item \textbf{FR+RJC+I/O}---Generate test cases based on FRs and RJC, {in addition to supplementary input-output pairs} that are extracted from test specifications.
    % We hope that LLMs can synthesize test cases based on these input-output pairs.
    
    \item \textbf{FR+RJC+TS}---Generate test cases based on FRs and RJC, {accompanied by test specifications}.
    We  anticipate that LLMs can implement test cases based on test specifications.
\end{itemize}
Afterwards, we compile generated test cases and run them against RJC.
Finally, we calculate \compo{} and \passo{}, and collect \clscov{}, \methcov{}, and \linecov{} from test executions.

\subsection{Prompt Templates}
\begin{figure}[!thb]
    \centering
    \subfloat[RQ1: DD+FR+DeM\label{fig:rq1-template}]{\includegraphics[width=0.327\linewidth]{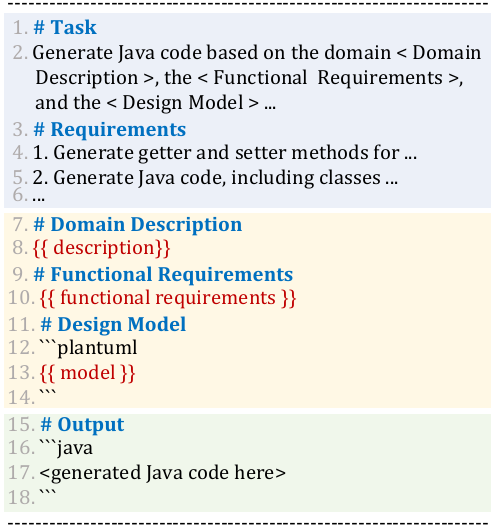}}
    \subfloat[RQ2: DD+FR \label{fig:rq2-template}]{\includegraphics[width=0.33\linewidth]{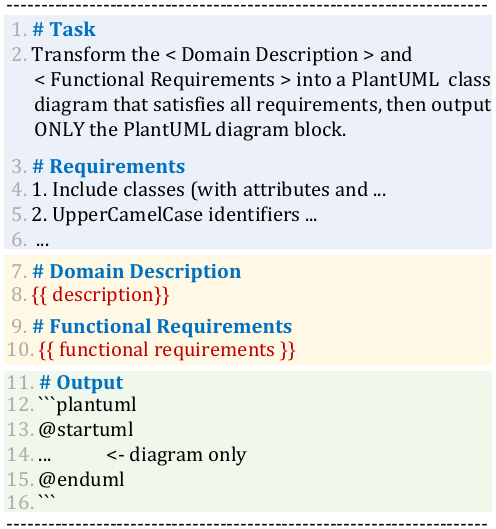}}
    \subfloat[RQ3: FR+RJC+TS\label{fig:rq3-template}]{\includegraphics[width=0.33\linewidth]{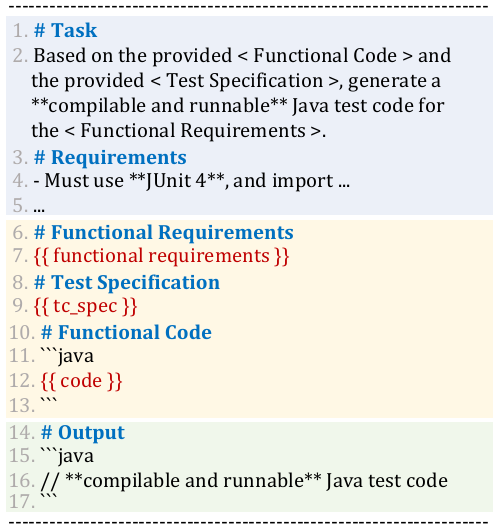}}
    \caption{Examples of prompt templates encoded in Jinjia2}\label{fig:prompt-template}
\end{figure}

Figure \ref{fig:prompt-template} illustrates the prompt templates used in the experiment. 
Each prompt template is structured into three distinct sections in Markdown format. 
The blue section presents the task description and step-by-step guidelines.
The yellow section includes a list of necessary context fragments, required by the task and the corresponding setting.
For example, in Figure \ref{fig:rq1-template}, the template for code generation includes a DD, FRs, and a DeM.
The green section specifies the expected output format.

\subsection{Selected LLMs}
Table \ref{tab:llms} presents the general information about the studied LLMs.
We select seven LLMs, including both open-source LLMs (two Qwen models and three DeepSeek models) and closed-source ones (two GPT models). 
For each category, we choose multiple models of different sizes, ranging from 7B to 671B.
Particularly, the selected DeepSeek models are reasoning models.
In our experiments, we set the temperature to be 0.7, leaving all other hyperparameters as default.

\begin{table}[!htb]
\caption{Studied LLMs}\label{tab:llms}
\resizebox{0.7\linewidth}{!}{
\begin{tabular}{lllll}
\hline
\textbf{Family} & \textbf{Model} & \textbf{Type} & \textbf{Size} & \textbf{Time} \\
\hline
\multirow{2}{*}{Qwen} & Qwen2.5-Coder-7B-instruct & Open-source & 7B & Sep, 2024 \\
                      & Qwen2.5-72B-Instruct & Open-source & 72B &  Sep, 2024 \\ \hline
\multirow{3}{*}{DeepSeek} & DeepSeek-R1-Distill-Qwen-7B & Open-source & 7B & Jan, 2025 \\
                          & DeepSeek-R1-Distill-Llama-70B & Open-source & 70B & Jan, 2025 \\
                          & DeepSeek-R1-671B  & Open-source & 671B & Jan, 2025 \\ \hline
\multirow{2}{*}{GPT} & GPT 3.5-Turbo-1106 & Closed-source & - & Nov, 2023 \\
                     & GPT 4o-Mini & Closed-source & - & Jul, 2024 \\
\hline
\end{tabular}
}
\end{table}

%% file: table/RQ1-setup.tex
\begin{table}[t!]
    \centering
    \renewcommand\arraystretch{1}
    \caption{Settings of RQ1 And RQ2}\label{tab:rq1-configs}
    \resizebox{0.75\linewidth}{!}{
    \begin{tabular}{l|lllll|l|l}
    \toprule
\multicolumn{1}{l|}{} & \multicolumn{5}{c|}{Input} & \multicolumn{1}{c}{\multirow{2}{*}{Output}} & \multicolumn{1}{c}{\multirow{2}{*}{Explanation}} \\ \cline{2-6}
\multicolumn{1}{l|}{ID} & \multicolumn{1}{c|}{DD} & \multicolumn{1}{c|}{FR} & \multicolumn{1}{c|}{MN} & \multicolumn{1}{c|}{DM} & \multicolumn{1}{c|}{CK} & \multicolumn{1}{c}{} & \multicolumn{1}{c}{} \\
\midrule
\multicolumn{8}{c}{\textbf{Settings for RQ1}} \\
\midrule
1 & Y & Y &  &  &  & Code Generation & No design provided \\
2 & Y & Y & Y &  &  & Code Generation & Only method names provided \\
3 & Y & Y &  & Y &  & Code Generation & Only high-level design provided \\
4 & Y & Y &  &  & Y & Code Generation & Only low-level design provided \\
5 & Y & Y &  & Y & Y & Code Generation & Both design levels provide \\
\midrule
\multicolumn{8}{c}{\textbf{Settings for RQ2}} \\
\midrule
6 & Y &  &  &  &  & Design Generation & \begin{tabular}[c]{@{}l@{}}Generate domain model \\ without operations\end{tabular} \\
7 & Y & Y &  &  &  & Design Generation & \begin{tabular}[c]{@{}l@{}}Generate design model \\ with operations\end{tabular} \\
8 & Y & Y &  &  &  & Design Generation & \begin{tabular}[c]{@{}l@{}}Generate domain model first,\\ then generate design model\end{tabular}\\
\bottomrule
\multicolumn{8}{l}{DD: Domain Description. FR: Functional Requirement. MN: Method Name.}\\\multicolumn{8}{l}{DM: Design Model. CS: Code Skeleton.}
\end{tabular}
}
\end{table}

%% file: sections/results.tex
\section{Results}
\subsection{RQ1: Design-aware code generation}

Table \ref{tab:rq1-overall-results} presents a structured comparison of performance metrics across multiple LLMs under different experimental settings. 
It is organized into two main metric categories: ‌\compk‌ and ‌\passk‌, each measured at three sizes (k=1, 3, 5). 
The rows are grouped by LLM variants (e.g., DeepSeek-R1, Qwen2.5, GPT series), with each model further divided into five experimental settings (DD+FR and its combinations with MN, DeM, JCS). 
The first column identifies the LLM, the second specifies the setting, and the subsequent columns display metric results. 
% font and color

\begin{table}[!htb]
    \centering
    \caption{Performance comparison of design-aware code generation}\label{tab:rq1-overall-results}
    \resizebox{0.7\linewidth}{!}{
    \begin{tabular}{l|l|rrr|rrr|rr}
    % {p{9em}|p{70pt}|p{10pt}p{10pt}p{10pt}|p{10pt}p{10pt}p{10pt}|p{25pt}p{25pt}}
    \hline
        \multirow{2}{*}{\textbf{LLM}} & \multirow{2}{*}{\textbf{Setting}} &  \multicolumn{3}{c|}{\textbf{\compk}}  &  \multicolumn{3}{|c}{\textbf{\passk}} &  \multicolumn{2}{|c}{\textbf{Match}}\\
        \cline{3---10}
         &  & \textbf{k=1} &\textbf{k=3} &\textbf{k=5} & \textbf{k=1} & \textbf{k=3} & \textbf{k=5} & \textbf{Class} & \textbf{Method} \\ \hline
\multirow{5}{*}{{\parbox{8em}{DeepSeek-R1-671B}}} & DD+FR & 0.84 & 0.99 & 1.00 & 0.00 & 0.00 & 0.00 & 0.67 & 0.37 \\
 & DD+FR+MN & 0.88 & 1.00 & 1.00 & 0.00 & 0.00 & 0.00 & 0.73& 0.91 \\
 & DD+FR+DeM & 0.86 & 0.99 & 1.00 & 0.03 & 0.06 & 0.06 & 0.98 & 0.88\\
 & DD+FR+JCS & \textbf{1.00} & \textbf{1.00} & \textbf{1.0} & \textbf{0.54} & \textbf{0.71} & \textbf{0.76} & \textbf{1.0}& \textbf{0.99}\\
 & DD+FR+DeM+JCS & 0.99 & \textbf{1.0} & \textbf{1.0} & 0.50 & 0.68 & 0.73 &\textbf{1.0} & \textbf{0.99}\\ \hline
\multirow{5}{*}{{\parbox{8em}{DeepSeek-R1-Distill-Llama-70B}}} & DD+FR & 0.67 & 0.95 & 0.99 & 0.00 & 0.00 & 0.00 & 0.64 & 0.39\\
 & DD+FR+MN & 0.75 & 0.98 & \textbf{1.0} & 0.00 & 0.00 & 0.00 & 0.70 & 0.86\\
 & DD+FR+DeM & 0.63 & 0.94 & 0.99 & 0.00 & 0.01 & 0.02 & 0.96 & 0.88\\
 & DD+FR+JCS & 0.93 & 0.99 & \textbf{1.0} & 0.39 & 0.54 & 0.58 & 0.98 & 0.96 \\
 & DD+FR+DeM+JCS & 0.92 & 0.99 & \textbf{1.0} & 0.46 & 0.66 & 0.73 & 0.99 & 0.98 \\ \hline
\multirow{5}{*}{{\parbox{8em}{DeepSeek-R1-Distill-Qwen-7B}}} & DD+FR & 0.06 & 0.17 & 0.26 & 0.00 & 0.00 & 0.00 & 0.49 & 0.27 \\
 & DD+FR+MN & 0.03 & 0.08 & 0.12 & 0.00 & 0.00 & 0.00 & 0.50 & 0.66\\
 & DD+FR+DeM & 0.12 & 0.31 & 0.46 & 0.00 & 0.01 & 0.02 & 0.81 & 0.76\\
 & DD+FR+JCS & 0.01 & 0.04 & 0.06 & 0.00 & 0.01 & 0.02 & 0.52 & 0.79\\
 & DD+FR+DeM+JCS & 0.01 & 0.04 & 0.07 & 0.00 & 0.00 & 0.00 &0.53 & 0.83 \\ \hline
\multirow{5}{*}{{\parbox{8em}{Qwen2.5-72B-Instruct}}} & DD+FR & 0.76 & 0.94 & 0.96 & 0.00 & 0.00 & 0.00 & 0.56 & 0.37\\
 & DD+FR+MN & 0.68 & 0.88 & 0.92 & 0.00 & 0.00 & 0.00 &0.67 &0.91\\
 & DD+FR+DeM & 0.73 & 0.93 & 0.98 & 0.02 & 0.05 & 0.06 &0.96 & 0.89 \\
 & DD+FR+JCS & 0.88 & 0.96 & 0.98 & 0.29 & 0.39 & 0.42 & \textbf{1.0} & \textbf{0.99}\\
 & DD+FR+DeM+JCS & 0.86 & 0.96 & 0.98 & 0.32 & 0.44 & 0.48 & \textbf{1.0} & \textbf{0.99} \\ \hline
\multirow{5}{*}{{\parbox{8em}{Qwen2.5-Coder-7B-Instruct}}} & DD+FR & 0.41 & 0.73 & 0.84 & 0.00 & 0.00 & 0.00 &0.62 & 0.37\\
 & DD+FR+MN & 0.50 & 0.82 & 0.94 & 0.00 & 0.00 & 0.00 & 0.62 & 0.86\\
 & DD+FR+DeM & 0.30 & 0.57 & 0.68 & 0.01 & 0.02 & 0.03 &0.96&0.91 \\
 & DD+FR+JCS & 0.63 & 0.85 & 0.92 & 0.21 & 0.33 & 0.37&0.99&0.98 \\
 & DD+FR+DeM+JCS & 0.73 & 0.91 & 0.95 & 0.21 & 0.33 & 0.38 &\textbf{1.0}&\textbf{0.99}\\ \hline
\multirow{5}{*}{{\parbox{8em}{GPT~4o-Mini}}} & DD+FR & 0.81 & 0.98 & \textbf{1.0} & 0.00 & 0.00 & 0.00 &0.61&0.41 \\
 & DD+FR+MN & 0.87 & 0.99 & \textbf{1.0} & 0.00 & 0.00 & 0.00  &0.61&0.84\\
 & DD+FR+DeM & 0.68 & 0.85 & 0.89 & 0.02 & 0.04 & 0.05 &0.99&0.93\\
 & DD+FR+JCS & 0.93 & 0.97 & 0.97 & 0.30 & 0.41 & 0.47&\textbf{1.0}&\textbf{0.99} \\
 & DD+FR+DeM+JCS & 0.93 & 0.99 & \textbf{1.0} & 0.33 & 0.43 & 0.47&\textbf{1.0}&\textbf{0.99} \\ \hline
\multirow{5}{*}{{\parbox{8em}{GPT~3.5-Turbo-1106}}} & DD+FR & 0.18 & 0.38 & 0.48 & 0.00 & 0.00 & 0.00&0.53&0.32 \\
 & DD+FR+MN & 0.88 & 0.99 & \textbf{1.0} & 0.00 & 0.00 & 0.00&0.62&0.84 \\
 & DD+FR+DeM & 0.40 & 0.66 & 0.77 & 0.00 & 0.00 & 0.00&0.97&0.89 \\
 & DD+FR+JCS & 0.72 & 0.91 & 0.96 & 0.20 & 0.36 & 0.43&0.80&0.96 \\
 & DD+FR+DeM+JCS & 0.58 & 0.84 & 0.90 & 0.16 & 0.28 & 0.33&0.69&0.94 \\ \hline
    \end{tabular}
    }
\end{table}

First, there is a pronounced and universal performance gap between \compk and \passk scores across all models and settings. 
In most cases, Compilation@5 scores are higher than 0.6, while most Pass@5 scores are lower than 0.5.
\textit{Achieving compilation is a much easier task for LLMs than generating functionally correct code that passes all test cases.} 
For instance,  DeepSeek-R1-671B achieves near-perfect Compilation@5 scores (1.00) under all settings, yet its Pass@5 scores range from an abysmal 0.00 to a more respectable 0.76.
Compared to other LLMs, DeepSeek-R1-Distill-Qwen-7B shows a different trend---its \compk scores are much lower ($\leq 0.46$).

Secondly, the choice of prompting strategy has a dramatic and disproportionate impact on performance. 
The addition of Java code skeleton (+JCS) emerges as the single most influential factor for improving correctness. 
With the exception of the smallest LLMs, the +JCS settings resulted in a massive leap in \passk scores, often from 0.30 to 0.76. 
Conversely, other settings like adding method names (+MN) or design models (DeM) alone show minimal to no positive effect on \passk, though they sometimes offer modest gains in \compk.
% Based on the observation, we derive Finding 1.

\begin{finding}
    \findingtitle{1}
    LLMs excel at managing low-level designs but struggle to understand high-level designs and to simultaneously perform software design and code generation.
\end{finding}

The low \passk scores under the settings without JCS are mainly caused by the missing and mismatching classes and methods that are required and invoked in test cases.
In the DD+FR setting, because LLMs are permitted to freely design, the code produced often deviates from the structure of the reference code (RJC) we supplied, leading to the failure of nearly all reference test cases. This is also confirmed by low \classmatch and \methodmatch scores.
The result is \textit{anticipated} since, when dealing with the same functionality, the LLMs might select alternative names and signatures.

Notably, even when we supply (partial) high-level designs, such as in DD+FR+MN and DD+FR+DeM settings, the performance of LLMs remains inadequate---the pass rate does not surpass 0.06. 
Upon an inspection of the code produced by LLMs and their corresponding test execution logs, it was observed that under both settings, LLMs still fail to comprehend and adhere to the given design. They frequently made the following design-related errors.
\begin{itemize}
    \item \textbf{Violation of given designs}\quad Despite the explicit provision of designs in the input, LLMs continue to struggle with adhering strictly to these designs when generating code.
    Table \ref{tab:rq1-overall-results} indicates that if \methodmatch scores fall below 0.95, the \passk scores quickly decrease to zero, since \textit{any single class/method mismatch will result in the generated code failing the tests}.
    In Figure~\ref{fig:rq1-error1-1}, an example produced by DeepSeek-R1-Distill-Llama-70B within the DD+FR+DeM configuration is illustrated. 
    The design model (Figure~\ref{fig:rq1-error1-1}-\ding{172}) adopts the role object pattern to allow a \texttt{User} to own multiple \texttt{UserRole}s, including \texttt{Author}, \texttt{Reviewer}, and \texttt{CoChair}, simultaneously.
    Conversely, in the generated Java code (GJC) depicted in Figure~\ref{fig:rq1-error1-1}-\ding{173}, DeepSeek-R1-Distill-Llama-70B modified the provided design so that \texttt{Author}, \texttt{Reviewer}, and \texttt{CoChair} subclass \texttt{User} directly.
    In this implementation, a \texttt{User} can only be an \texttt{Author}, a \texttt{Reviewer}, or a \texttt{CoChair}.
    \begin{figure}[!htb]
        \centering
        \includegraphics[scale=0.6]{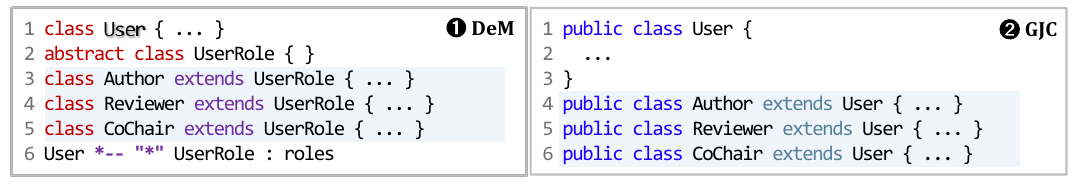}
        \caption{Incorrect inheritances}
        \label{fig:rq1-error1-1}
    \end{figure} 

    Figure~\ref{fig:rq1-error1-2} shows an example generated by DeepSeek-R1-671B under the DD+FR+DeM setting. 
    The problem illustrated in Figure~\ref{fig:rq1-error1-2}-\ding{172} arises at line 5 due to the missing method \texttt{calculateTotalPrice()}.
    The design model, as shown in Figure~\ref{fig:rq1-error1-2}-\ding{173}, distinctly specifies this operation in the class \textit{Artist}.
    The generated code does implement this operation but with a different name \texttt{calculateTotalArtwork\-Price()}, as shown in Figure~\ref{fig:rq1-error1-2}-\ding{174}. 
    This variation in method naming leads to execution failure.
    % figure for example 2
    \begin{figure}[!htb]
        \centering
        \includegraphics[width=1\linewidth]{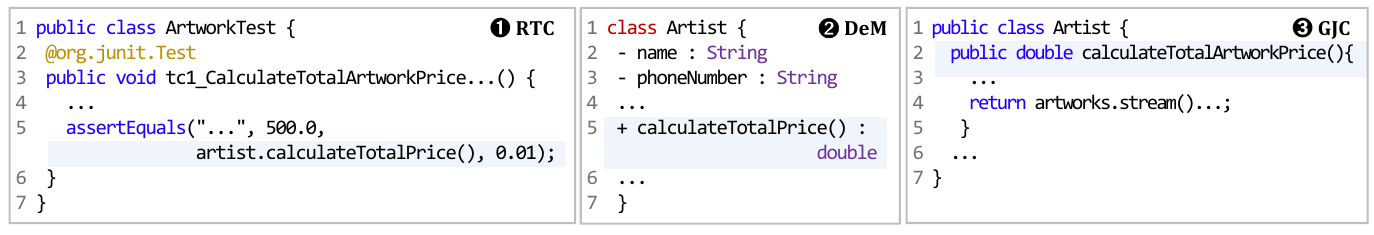}
        \caption{Variation in method names}
        \label{fig:rq1-error1-2}
    \end{figure} 

    Figure~\ref{fig:rq1-error1-3} presents an alternative instance produced by DeepSeek-R1-671B within the DD+FR+DeM configuration. 
    The execution encounters an error at line 6 in \ref{fig:rq1-error1-3}-\ding{172} caused by a type mismatch of the parameter in the method \texttt{checkOrderFeasible(currentDate)}. 
    The design model, depicted in Figure~\ref{fig:rq1-error1-3}-\ding{173}, explicitly defines that the operation \texttt{checkOrderFeasible()} of the \textit{Order} class has a parameter of type \texttt{Date}. 
    In contrast, the generated code implements it using a Java method \texttt{checkOrderFeasible(LocalDateTime currentDate)}, as indicated in Figure~\ref{fig:rq1-error1-3}-\ding{174}, with a different parameter type \texttt{LocalDateTime}. 

    \begin{figure}[!htb]
        \centering
        \includegraphics[width=1\linewidth]{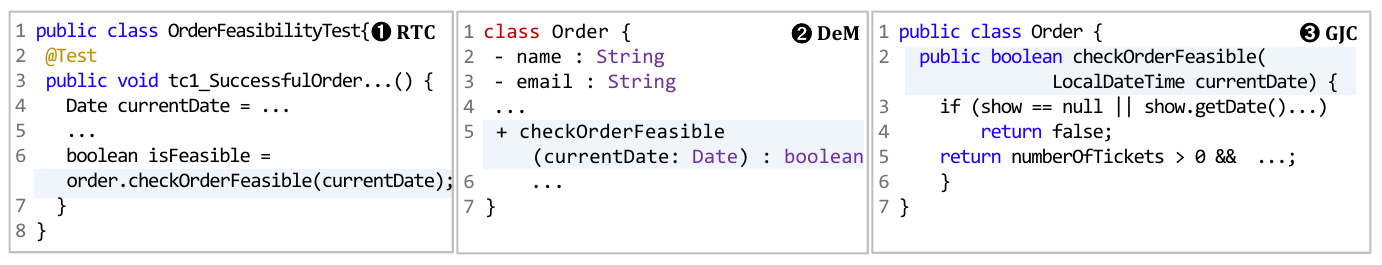}
        \caption{Inconsistent parameter types}
        \label{fig:rq1-error1-3}
    \end{figure}

    \item \textbf{Lacking the knowledge of translating design into code}  
    LLMs usually lack the specific programming knowledge to implement software design correctly, leading to compilation and runtime errors.
    Figure~\ref{fig:rq1-error2-1} shows an example generated by DeepSeek-R1-Distill-Llama-70B under the DD+FR+DeM setting. 
    The compilation is unsuccessful at line 6 in Figure~\ref{fig:rq1-error2-1}-\ding{172} because of the absence of the method \texttt{getCourses()}. 
    In the design model depicted in Figure~\ref{fig:rq1-error2-1}-\ding{173}, the class \texttt{AcademicProgram} contains an attribute \texttt{courses}. 
    It is a general principle when translating a class diagram into code to define a getter method for each attribute included in the model. 
    Despite explicitly prompting LLMs to generate getters and setters, as detailed in Figure~\ref{fig:rq1-template}, DeepSeek-R1-Distill-Llama-70B did not adhere to our instructions nor follow the general rule.

    \begin{figure}[!htb]
        \centering
        \includegraphics[scale=0.6]{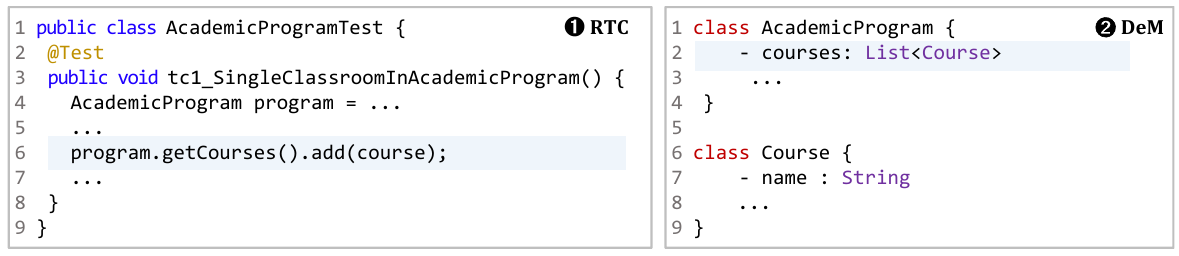}
        \caption{Missing getter methods}
        \label{fig:rq1-error2-1}
    \end{figure} 
    
   LLMs struggle with importing data type classes. 
   As seen in Figure~\ref{fig:rq1-error2-2}-\ding{172}, generated by DeepSeek-R1-Distill-Llama-70B under the DD+FR+DeM setting, a compilation error at line 5 occurs due to missing \texttt{import java.util.Date}. 
   The \texttt{Date} type, used in the model shown in Figure~\ref{fig:rq1-error2-2}-\ding{173}, is a basic type in many modeling tools like EMF/Ecore. When converting an Ecore model to Java, \texttt{java.util.Date} must be imported explicitly if it occurs in the model. LLMs do not automatically recognize this necessity when generating code from DeMs but instead depend on runtime inference to determine the need for correct import statements dynamically.
    \begin{figure}[!htb]
        \centering
        \includegraphics[scale=0.6]{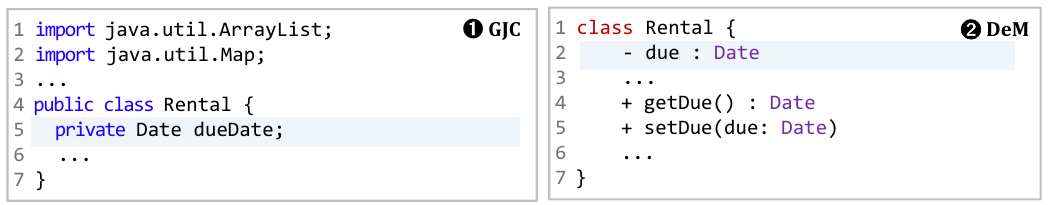}
        \caption{Missing import statements for data types}
        \label{fig:rq1-error2-2}
    \end{figure} 
     
    Figure~\ref{fig:rq1-error2-3} presents another instance generated by DeepSeek-R1-671B using the DD+FR+DeM setting. In execution, the assertion on line 9 of Figure~\ref{fig:rq1-error2-3}-\ding{172} failed, expecting \texttt{TotalAmount} to be 5500, but it was set to 11000. Upon examining the code, we observed that in the \texttt{setCustomer} method of the \texttt{Application} class, DeepSeek inserted lines 5--7 in Figure~\ref{fig:rq1-error2-3}-\ding{173}, which appended the \texttt{Application} to \texttt{customer}'s \texttt{applications} list. However, the design model in Figure~\ref{fig:rq1-error2-3}-\ding{174} explicitly shows that \texttt{customer}'s \texttt{applications} and \texttt{Application}'s \texttt{company} are distinct and can be modified \textit{independently}. Misinterpretation by DeepSeek R1 caused this erroneous code due to hallucination, mistakenly linking two unrelated associations. When the test code in Figure~\ref{fig:rq1-error2-3}-\ding{172} executed lines 6 and 7, \texttt{app1} was redundantly added to \texttt{customer} twice, inflating \texttt{totalAmount} and triggering the assertion failure.

    \begin{figure}[!htb]
        \centering
        \includegraphics[scale=0.6]{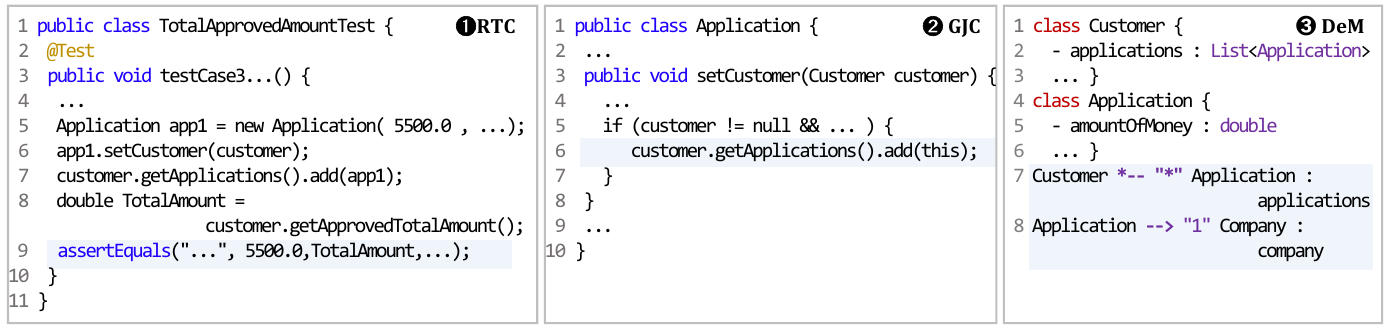}
        \caption{Misunderstanding the semantics of associations}
        \label{fig:rq1-error2-3}
    \end{figure}

\end{itemize}

Given that the UML class diagram is the central model for OO design, a comparison among the settings with and without +DeM is in order. 
For \compk, most LLMs, with the exception of DeepSeek-R1-Distill-Qwen-7B and Qwen2.5-72B-Instruct, exhibit inferior performance in the DD+FR+DeM setting as opposed to the DD+FR+MN setting. 
When compared to DD+FR, Qwen2.5-Coder-7B-Instruct and GPT 4o-Mini also show reduced performance in the DD+FR+DeM setting. 
In evaluating the DD+FR+JCS relative to the DD+FR+DeM+JCS settings, the introduction of DeMs exerts only a marginal effect on improving code syntax accuracy, with most LLMs reflecting minor advancements, and some, notably GPT 3.5-Turbo-1106, experiencing a decrease in \compk scores. 
Conversely, regarding \passk, models such as DeepSeek-R1-Distill-Llama-70B, Qwen2.5-72B-Instruct, and Qwen2.5-Coder-7B-Instruct enhance the functional correctness of generated code with DeM, while other LLMs, like GPT 3.5-Turbo-1106, show a reduction in \passk scores in the DD+FR+DeM+JCS setting.
It is important to note that despite DeMs being encoded in PlantUML, which bears similarity to Java code, LLMs appear to lack the capability to generate code grounded on these designs (with \methodmatch<0.93).
Based on these observations, Finding 2 can be derived.

\begin{finding}
    \findingtitle{2}
    Conventionally, design models serve as crucial artifacts that enhance communication among developers. However, LLMs have difficulty adhering to the design knowledge and constraints outlined in these models during code generation.
\end{finding}

We hypothesize that design models can hinder code generation in LLMs for two primary reasons. First, LLMs may have limited exposure to PlantUML syntax, potentially impeding their interpretation of models written in this language. 
That said, PlantUML class diagram syntax shares notable similarities with object-oriented programming languages like Java and Python, so their generalization capabilities should, in theory, allow for some comprehension. 
The second, more critical reason is that LLMs struggle to translate design into implementation, as they lack the conventional and implicit knowledge and constraints governing this transformation. 
For example, implementing class attributes often requires the creation of getter and setter methods, and associations between classes usually imply potential method calls for interoperation. 
LLMs are often unaware of or unable to apply these rules. 
Consequently, the provided design model acts as interference rather than a guide.

Conversely, supplying a code skeleton (+JCS) notably enhances the performance of LLMs. 
This enhancement occurs because JCS encompasses the details for translating design into code. 
Despite the expansion of context, LLMs can focus on completing unimplemented methods without the added complexity of addressing designs, thereby increasing their problem-solving efficacy.

Thirdly, model size has a significant impact on performance, with larger models (e.g., DeepSeek-R1-671B, Qwen2.5-72B, GPT~4o-Mini) consistently delivering superior results compared to smaller ones (e.g., DeepSeek-R1-Distill-Qwen-7B, Qwen2.5-Coder-7B). This trend underscores how the capacity of a model is crucial for effective code generation in specific tasks. Given identical input conditions, the hierarchy within the DeepSeek series is as follows: DeepSeek-R1-671B is the top performer, followed by DeepSeek-R1-Distill-Llama-70B, and then DeepSeek-R1-Distill-Qwen-7B. In the Qwen series, Qwen2.5-72B-Instruct outranks Qwen2.5-Coder-7B-Instruct. At the 70B parameter level, DeepSeek-R1-Distill-Llamma-70B has a slight edge over Qwen2.5-72B-Instruct. In contrast, at the 7B parameter level, Qwen2.5-coder, which specializes in coding tasks, significantly outperforms the general-purpose DeepSeek-R1-7B. Among closed-source models, GPT~4o-Mini consistently surpasses GPT~3.5-Turbo-1106 under a variety of input configurations.
Moreover, aside from the two 7B models, open-source models demonstrate comparable performance to the closed-source models we have evaluated in tasks related to design-aware code generation.

\subsection{RQ2: Object-oriented design}

Table \ref{tab:oom-result} presents the effectiveness of LLMs in creating object-oriented design structures (i.e., class diagrams).
The first column enumerates the LLMs used; the second column describes the various generation configurations; columns 3--17 provide the precision, recall, and F1 scores for Classes, Attributes, Operations, Inheritances, and Associations in the generated class diagrams relative to the reference models.
Because LLMs are not required to generate operations in DD settings, the corresponding metric values are marked as N/A.

\begin{table}[t!]
\centering
\caption{Results of object-oriented design model generation}\label{tab:oom-result}
\renewcommand\arraystretch{1.06}
\resizebox{1.0\linewidth}{!}{
\begin{tabular}{l|l|rrr|rrr|rrr|rrr|rrr}
% {p{6em}|l|p{10pt}p{10pt}p{10pt}|p{10pt}p{10pt}p{10pt}|p{10pt}p{10pt}p{10pt}|p{10pt}p{10pt}p{10pt}|p{10pt}p{10pt}p{10pt}}
\toprule
\multirow{2}{*}{\textbf{LLM}} & \multirow{2}{*}{\textbf{Setting}} & \multicolumn{3}{c|}{\textbf{Class}} & \multicolumn{3}{c|}{\textbf{Attribute}} & \multicolumn{3}{c|}{\textbf{Operation}} & \multicolumn{3}{c|}{\textbf{Inheritance}} & \multicolumn{3}{c}{\textbf{Association}} \\
& & \preci & \recal & \fone & \preci & \recal & \fone & \preci & \recal & \fone & \preci & \recal & \fone & \preci & \recal & \fone \\
\midrule
\multirow{3}{*}{\parbox{6em}{DeepSeek-R1-671B}}                 & DD        & 0.80&	0.87&	0.83&	0.65&	0.72&	0.68&	N/A&	N/A&	N/A&	0.65&	0.84&	0.73&	0.63&	0.60&	0.61  \\
& DD+FR     &\textbf{0.82}&	0.88&	\textbf{0.85}&	\textbf{0.66}&	\textbf{0.74}&	\textbf{0.70}&	0.55&	0.58&	0.56&	\textbf{0.76}&	\textbf{0.91}&	\textbf{0.83}&	0.73&	0.70&	\textbf{0.72} \\
& DD+FR*   & \textbf{0.82}&	0.88&	\textbf{0.85}&	0.62&	0.73&	0.67&	0.58&	0.59&	\textbf{0.59}&	0.73&	0.88&	0.80&	0.64&	0.64&	0.64  \\ \hline
\multirow{3}{*}{\parbox{6em}{DeepSeek-R1-Distill-Llama-70B}}    & DD        & 0.77&	0.84&	0.80&	0.60&	0.72&	0.66&	N/A&	N/A&	N/A&	0.66&	0.81&	0.73&	0.63&	0.63&	0.63 \\
& DD+FR      & 0.79&	\textbf{0.89}&	0.84&	0.64&	0.73&	0.68&	0.49&	0.52&	0.50&	0.74&	0.79&	0.76&	0.67&	\textbf{0.72}&	0.69   \\
& DD+FR*   &0.77&	0.87&	0.81&	0.52&	0.71&	0.60&	\textbf{0.65}&	0.55&	\textbf{0.59}&	0.69&	0.81&	0.74&	0.71&	0.69&	0.70  \\ \hline
\multirow{3}{*}{\parbox{6em}{DeepSeek-R1-Distill-Qwen-7B}}      & DD        &0.73&	0.67&	0.70&	0.43&	0.50&	0.46&	N/A&	N/A&	N/A&	0.10&	0.12&	0.11&	0.34&	0.33&	0.34  \\
& DD+FR     &0.77&	0.79&	0.78&	0.70&	0.68&	0.69&	0.58&	0.52&	0.55&	0.50&	0.70&	0.58&	0.71&	0.57&	0.63    \\
& DD+FR*    &0.78&	0.68&	0.72&	0.54&	0.59&	0.56&	0.58&	0.50&	0.53&	0.18&	0.14&	0.16&	0.54&	0.42&	0.47 \\ \hline
\multirow{3}{*}{\parbox{6em}{Qwen2.5-72B-Instruct}}             & DD        &0.69&	0.88&	0.77&	0.49&	0.70&	0.57&	N/A&	N/A&	N/A&	0.42&	0.74&	0.54&	0.56&	0.60&	0.58 \\
& DD+FR     &0.75&	0.82&	0.79&	0.63&	0.72&	0.67&	0.48&	0.59&	0.53&	0.48&	0.64&	0.55&	0.64&	0.56&	0.60 \\
& DD+FR*    &0.70&	0.87&	0.78&	0.52&	0.71&	0.60&	0.47&	\textbf{0.61}&	0.53&	0.44&	0.74&	0.55&	0.64&	0.71&	0.67 \\ \hline
\multirow{3}{*}{\parbox{6em}{Qwen2.5-Coder-7B-Instruct}}        & DD        &0.68&	0.72&	0.70&	0.50&	0.55&	0.53&	N/A&	N/A&	N/A&	0.24&	0.33&	0.27&	0.43&	0.44&	0.43 \\
& DD+FR     &0.76&	0.80&	0.78&	0.69&	0.70&	0.69&	0.39&	0.45&	0.42&	0.53&	0.60&	0.57&	0.63&	0.58&	0.60  \\
& DD+FR*   &0.77&	0.76&	0.76&	0.53&	0.60&	0.56&	0.37&	0.51&	0.43&	0.44&	0.33&	0.37&	0.55&	0.46&	0.50  \\ \hline
\multirow{3}{*}{\parbox{6em}{GPT 4o-Mini}}                      & DD        & 0.76&	0.79&	0.77&	0.60&	0.66&	0.63&	N/A&	N/A&	N/A&	0.52&	0.74&	0.62&	0.67&	0.50&	0.57  \\
& DD+FR       &0.75&	0.78&	0.77&	0.65&	0.70&	0.67&	0.53&	0.53&	0.53&	0.67&	0.79&	0.72&	0.71&	0.51&	0.59  \\
& DD+FR*   &0.78&	0.79&	0.78&	0.65&	0.73&	0.69&	0.53&	0.52&	0.53&	0.58&	0.79&	0.67&	0.70&	0.50&	0.58 \\ \hline
\multirow{3}{*}{\parbox{6em}{GPT 3.5-Turbo-1106}}               & DD        &0.78&	0.69&	0.73&	0.61&	0.59&	0.60&	N/A&	N/A&	N/A&	0.33&	0.40&	0.36&	0.45&	0.35&	0.39  \\
& DD+FR     &0.77&	0.76&	0.76&	0.65&	0.68&	0.67&	0.57&	0.53&	0.55&	0.52&	0.70&	0.59&	\textbf{0.77}&	0.53&	0.63 \\
& DD+FR*   & 0.78&	0.72&	0.75&	0.63&	0.59&	0.61&	0.54&	0.57&	0.55&	0.31&	0.47&	0.37&	0.56&	0.43&	0.49  \\ \hline
\end{tabular}
}
\end{table}

From Table \ref{tab:oom-result}, it is clear that all LLMs excel at object and class definition, as indicated by the F1 scores for classes, which consistently surpass 0.7 in each context, reaching a peak of 0.85.
LLMs show reduced ability with structural features like attributes and associations compared to classes. Attribute F1 scores are between 0.46 and 0.7 (median: 0.67), while association scores range from 0.34 to 0.72 (median: 0.6). LLMs handle attributes better than associations. Inheritance handling varies: DeepSeek-R1-671B and DeepSeek-R1-Distill-Llama-70B scores exceed 0.7, but DeepSeek-R1-Distill-Qwen-7B falls below 0.2. Operation scores range from 0.42 to 0.59 (median: 0.53), exceeding 0.5 except for Qwen2.5-Coder-7B-Instruct. Though LLMs identify classes well, they struggle with design operations and class relationships.

Within the same family of LLMs, the parameter size is a key determinant of the quality of a generated design model. 
The 671B version of DeepSeek surpasses the capabilities of its 70B and 7B counterparts significantly, while the 72B variant of Qwen is superior to its 7B version. In the context of closed-source models, GPT 4o-Mini exhibits enhanced performance compared to GPT 3.5-Turbo-1106.
The impact of parameter size in LLMs on the creation of classes and attributes is slightly less significant than its effect on the generation of other types of elements. 
This suggests that LLMs exhibit a more stable understanding of the concepts of class and attribute.

\begin{finding}
    \findingtitle{3}
   
    LLMs are proficient and reliable at recognizing objects and classes, yet they demonstrate weaker abilities in defining operations and inter-class relationships, underscoring the need to enhance LLMs' design skills for managing systems with multiple interdependent modules.
\end{finding}

Evaluating different generation settings shows that integrating FR typically improves DeM generation quality, regardless of the LLM used. FRs describe business logic involving the objects mentioned in DD, thereby fortifying these objects' definitions and relationships, which allows LLMs to create higher-quality DeMs. In less capable LLMs (e.g., 7B-parameter models), FR incorporation results in a more comprehensive enhancement of DeM generation.

Comparing one-step (DD+FR) and two-step (DD+FR*) generation shows that the one-step approach modestly improves static feature generation like classes, attributes, associations, and inheritances in most LLMs, thus raising their F1 scores. For example, under DD+FR, GPT 3.5-Turbo-1106 saw F1 score increases of 0.01, 0.06, 0.14, and 0.22 for these features compared to DD+FR*. However, this method isn't as effective for operations and can even reduce scores for some LLMs. For instance, DeepSeek-R1-671B's operations F1 score fell by 0.03 under DD+FR. In LLMs with fewer parameters, the one-step setup enhances static features but has minor effects on dynamic ones. For example, with Qwen2.5-Coder-7B-Instruct, F1 scores for class, attribute, association, inheritance, and operations under DD+FR changed by +0.02, +0.13, +0.1, +0.2, and -0.01, respectively. Additionally, LLMs with strong reasoning, like DeepSeek series, perform better in two-step operation generation F1 scores, while those with weaker reasoning, like the Qwen series, maintain stable F1 scores for operations.

In class diagrams, static features (such as classes, attributes, inheritance, and associations) and dynamic features (like operations) are intertwined. Static features define the core data structures, limiting the data accessible for dynamic features. In return, the involvement of static features in dynamic processes highlights key system data, shaping their design. Examining DeepSeek R1's reasoning, the one-step generation approach enables LLMs to access crucial data for operations, potentially boosting F1 scores for static features by better identifying attributes and relationships. However, this approach relies heavily on DD and FR, leading to relaxed constraints and possible errors in modeling. Conversely, the two-step strategy asks LLMs to outline operations using existing domain models (class diagrams with static features), optimizing the utilization of extra information for creating dynamic features.

Manual review of generated design models revealed that class generation quality critically determines the accuracy and completeness of other parts. 
Incorrect class generation leads to incorrect or missing attributes, relationships, and operations, thereby degrading metrics for other model elements.
Besides, we also noted that LLMs commonly make the following types of mistakes:

\begin{itemize}
\item \textbf{Conceptual ambiguity} arises when LLMs misinterpret modeling concepts, resulting in incorrect or misleading models. For instance, LLMs frequently misrepresent messages between objects as associations. 
As illustrated in Figure \ref{fig:rq2-error-cases}-\ding{172}, according to DD, when a \texttt{CoChair} make a decision for a \texttt{Paper}, it sends a \textit{message} to the \texttt{Paper} to change its internal state.
However, GPT-4o-Mini incorrectly established an association on line 7 to represent this message.
Additionally, in Figure \ref{fig:rq2-error-cases}-\ding{173}, DeepSeek-R1-Distill-Llama-70B mistakenly created an association between the class \texttt{Paper} and an enumeration type \texttt{PaperType}, which should be defined as an attribute. % 去掉图2中第7行

\item \textbf{Overdesign} occurs when LLMs create superfluous or redundant structures. 
As depicted in Figure \ref{fig:rq2-error-cases}-\ding{174}, LLMs creates a parent class \texttt{Pet} with two sub-classes, \texttt{Dog} and \texttt{Cat}, while concurrently defining an enumeration type \texttt{PetType}. 

In this example, the inheritance hierarchy and the enumeration attribute are redundant, and it is necessary to retain only one of these structures.
In Figure \ref{fig:rq2-error-cases}-\ding{175}, line 8 illustrates an overdesign by Qwen2.5-72B-Instruct, as one can determine all \texttt{Participant}s of a \texttt{School} via the associations on lines 3–6.

\begin{figure}[!htb]
    \centering
    \includegraphics[scale=0.55]{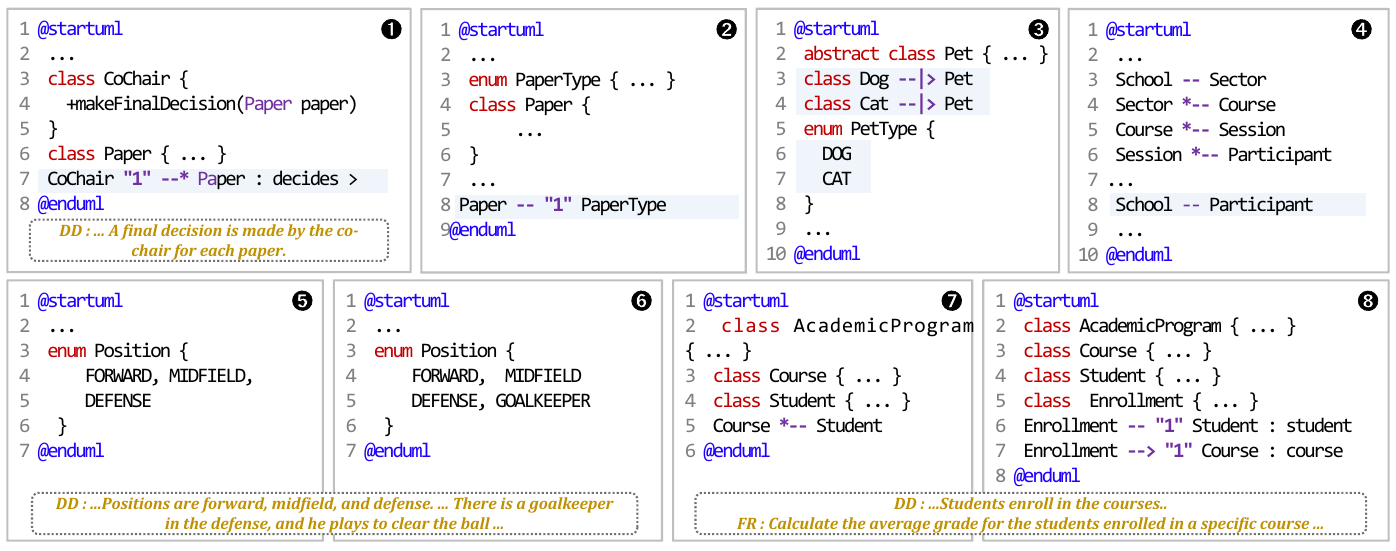}\\ 
    \caption{Error cases of model generation}
    \label{fig:rq2-error-cases}
\end{figure}

\item \textbf{Underdesign}, whereby LLMs frequently struggle to accurately deduce objects, attributes, and associations that are not explicitly indicated. 
For instance, in Figure \ref{fig:rq2-error-cases}-\ding{176}, three literals are listed in the generated model. However, Figure \ref{fig:rq2-error-cases}-\ding{177} shows our DeM incorporating an additional literal \texttt{GOALKEEPER}. 
Despite DD stating that "There is a goalkeeper in the defense," in a soccer team, a goalkeeper is different from general defense players because they can use their hands.
Furthermore, Figure \ref{fig:rq2-error-cases}-\ding{178}, produced by DeepSeek-R1-Distill-Llama-70B, omits a crucial class, namely \texttt{Enrollment}, which serves as a link between \texttt{Course} and \texttt{Student}, as demonstrated in Figure \ref{fig:rq2-error-cases}-\ding{179}. While DD does not explicitly reference enrollment objects, this class is crucial to document extra details, like course scores.

\item \textbf{Grammar issues} refer to syntactical inaccuracies in the generated model notation, which may prevent the model from being parsed or interpreted. 
Typical errors encompass omitted closing brackets ``\}'', particularly in class definitions, and incorrect usage of relationship symbols, such as ``\texttt{*-->}'', ``\texttt{<-+}'', ``\texttt{-->}'', ``\texttt{<..}'' for associations. These issues compromise the effective use of the resulting design models, as current modeling tools are unable to parse the generated models.
\end{itemize}

\subsection{RQ3: Acceptance test design and generation}

Table \ref{fig:rq3-results} presents the results of the design and generation of acceptance test cases.
The first column enumerates the LLMs used; the second column lists the various settings; columns 3--7 provide the values of \compo, \passo, \clscov, \methcov, and \linecov.

\begin{table}[!thb]
\caption{Results of acceptance test cases design and generation}\label{fig:rq3-results}
\scriptsize{
\begin{tabular}{p{8em}|l|cc|ccc}
\hline
\textbf{LLMs} & \textbf{Setting} & \textbf{\compo} & \textbf{\passo} & \textbf{\clscov} & \textbf{\methcov} & \textbf{\linecov} \\
\hline
\multirow{3}{*}{\parbox{8em}{DeepSeek-R1-671B}}             & FR+RJC     &\textbf{99\%}&95\%&	0.97&	0.71&	\textbf{0.80}\\
                                                            & FR+RJC+I/O &95\%&\textbf{99\%}&0.98&0.72&\textbf{0.80}\\
                                                            & FR+RJC+TS  &96\%&\textbf{99\%}&\textbf{0.99}&0.72&0.79\\ \hline
\multirow{3}{*}{\parbox{8em}{DeepSeek-R1-Distill-Llama-70B}}& FR+RJC     &87\%&91\%&0.96&0.70&0.78\\
                                                            & FR+RJC+I/O &87\%&94\%&0.97&0.70&0.77\\
                                                            & FR+RJC+TS  &86\%&94\%&0.98&0.71&0.77 \\ \hline
\multirow{3}{*}{\parbox{8em}{DeepSeek-R1-Distill-Qwen-7B}}  & FR+RJC     &2\%&11\%&0.14&0.08&0.08 \\
                                                            & FR+RJC+I/O &5\%&32\%&0.29&0.14&0.15\\
                                                            & FR+RJC+TS  &6\%&41\%&0.47&0.23&0.23\\ \hline
\multirow{3}{*}{\parbox{8em}{Qwen2.5-72B-Instruct}}         & FR+RJC     &84\%&80\%&0.97&0.72&0.78\\
                                                            & FR+RJC+I/O &79\%&83\%&0.97&\textbf{0.73}&0.79\\
                                                            & FR+RJC+TS  &76\%&89\%&0.98&\textbf{0.73}&0.78\\ \hline
\multirow{3}{*}{\parbox{8em}{Qwen2.5-Coder-7B-Instruct}}    & FR+RJC     &42\%&67\%&0.95&0.67&0.72\\
                                                            & FR+RJC+I/O &46\%&72\%&0.97&0.65&0.70\\
                                                            & FR+RJC+TS  &44\%&79\%&0.95&0.64&0.67\\ \hline
\multirow{3}{*}{\parbox{8em}{GPT 4o-Mini }}                 & FR+RJC     &99\%&82\%&0.96&0.69&0.77\\
                                                            & FR+RJC+I/O &98\%&86\%&0.97&0.69&0.77\\
                                                            & FR+RJC+TS  &97\%&91\%&0.98&0.71&0.79\\ \hline
\multirow{3}{*}{\parbox{8em}{GPT 3.5-Turbo-1106 }}          & FR+RJC     &96\%&79\%&0.96&0.71&0.78\\
                                                            & FR+RJC+I/O &98\%&84\%&0.98&0.69&0.77\\
                                                            & FR+RJC+TS  &90\%&86\%&0.98&0.71&0.78 \\ \hline
\end{tabular}
}
\end{table}

According to Table \ref{fig:rq3-results}, LLMs demonstrate robust proficiency in test case design and generation. 
The DeepSeek-R1-671B model showed superior performance among all evaluated LLMs, with more than 95\% of its test cases compiling without errors; furthermore, over 95\% of these compiled cases enabled the associated Java code to successfully pass the tests. 
Although DeepSeek-R1-Distill-Llama-70B exhibited a higher incidence of compilation errors, achieving a successful compilation rate of over 86\%, its test pass rate was still 91\%. 
In contrast, GPT 4o-Mini and GPT 3.5-Turbo-1106, while attaining high \compo scores, had notably lower \passo compared to DeepSeek-R1-671B and DeepSeek-R1-Distill-Llama-70B. 
The Qwen2.5-72B-Instruct model performed below GPT 4o-Mini, with both \compo and \passo scores over 76\% and 80\%, respectively.
In terms of code coverage, including \clscov, \methcov, and \linecov, DeepSeek-R1-671B, DeepSeek-R1-Distill-Llama-70B, Qwen2.5-72B-Instruct, GPT 4o-Mini, and GPT 3.5-Turbo-1106 demonstrate comparable performance. 
Remarkably, DeepSeek-R1-671B achieves class coverage scores reaching as high as 0.99 along with line coverage at 0.80. 
In contrast, Qwen2.5-72B-Instruct outperforms in method coverage, attaining a value of 0.73.
For a given family of LLMs, an increase in model parameters typically leads to enhanced performance in both designing and generating test cases.

For the majority of LLMs, the inclusion of input-output pairs (+I/O) and test specifications (+TS) tends to decrease the success rate of generating test cases that compile. 
This phenomenon may arise because the additional constraints imposed by extra information can disrupt the test case generation process, increasing the chance of syntactical errors. 
However, smaller models, such as DeepSeek-R1-Distill-Qwen-7B and Qwen2.5-Coder-7B-Instruct, actually see improved \compo performance with +I/O and +TS settings. Irrespective of the model used, employing +I/O and +TS settings consistently results in elevated \passo scores.
By comparing different settings, it is clear that LLMs are more likely to generate logically incorrect test cases when designing them autonomously. 
For example, the test cases generated autonomously by DeepSeek-R1-671B have a \passo score of 95\%, but after some test design is provided (+I/O or +TS), its \passo score increases to 99\%.
The more test designs we provided, the higher \passo scores LLMs achieved.

\begin{figure}[!thb]
    \centering
    \includegraphics[scale=0.45]{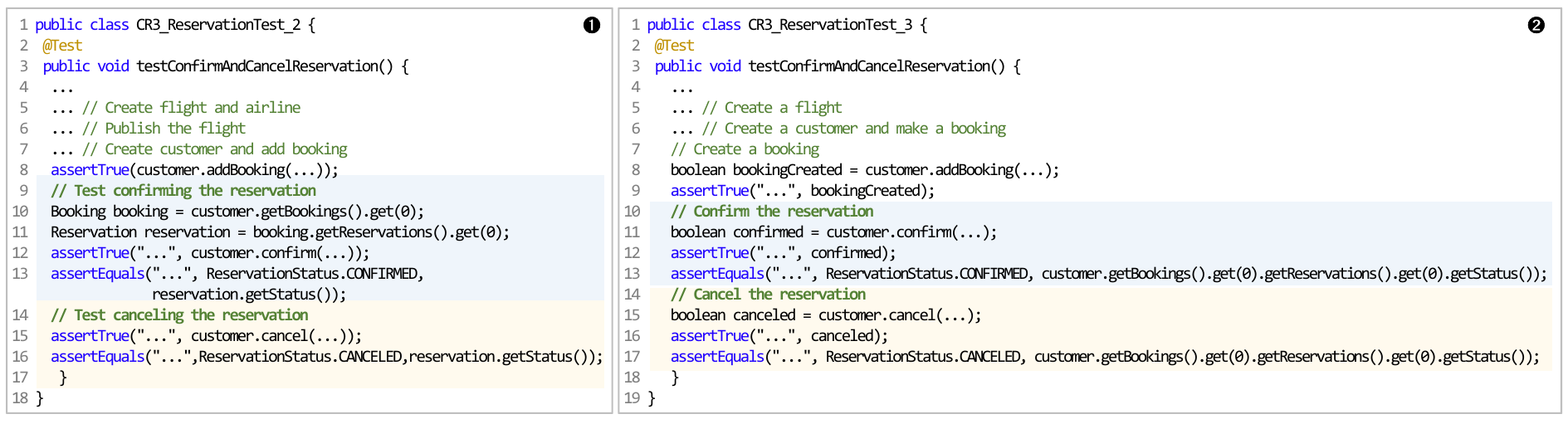}
    \caption{Example test cases generated by DeepSeek-R1-Distill-Llama-70B}
    \label{fig:rq3-error}
\end{figure}

Nevertheless, when considering code coverage metrics, supplying extra test designs does not significantly affect the outcomes. 
In comparison to the FR+RJC setting, the FR+RJC+TS setting results in a minor enhancement of class coverage, while its effect on method coverage and line coverage remains ambiguous. 
Notably, for Qwen2.5-Coder-7B-Instruct, +TS actually reduces line coverage from 0.72 to 0.67.
Thus, considering only coverage aspects, the quality of test cases autonomously designed by LLMs is fundamentally similar to those designed manually by humans.
\begin{finding}
    \findingtitle{4}
    Acceptance test cases crafted by LLMs based on functional requirements can attain equivalent code coverage to those devised by humans.
\end{finding}

We explored the test files generated by DeepSeek-R1-Distill-Llama-70B under the FR+RJC setting and observed the following problems regarding test design.

% example generated by DeepSeek-R1-Distill-Llama-70b 
\begin{itemize}
    \item 
    We consistently implement one test case per test method.
    However, LLMs often create multiple test cases within one single test method (e.g., Figure \ref{fig:rq3-error}-\ding{172} and Figure \ref{fig:rq3-error}-\ding{173}), even though we restricted the number of generated test cases. 
    Such a test design may negatively affect test isolation, readability, and maintainability, leading to a challenging debugging process. 
    \item LLMs frequently generate similar test cases. 
    For example, 
    the test cases in Figure \ref{fig:rq3-error}-\ding{172} are equivalent to those in Figure \ref{fig:rq3-error}-\ding{173}.
    
    \item LLMs may also miss critical edge cases. As shown in Figure \ref{fig:rq3-error}-\ding{172} and Figure \ref{fig:rq3-error}-\ding{173}, the LLM tested the \texttt{CONFIRMED} and \texttt{CANCELED} scenarios but consistently overlooked the \texttt{PENDING} state.
\end{itemize}

\begin{finding}
    \findingtitle{5}
    LLMs currently lack the ability to devise the same level of concise and varied acceptance test cases that humans are capable of.
\end{finding}

\subsection{Threats to Validity}
This subsection briefly discusses the major threats to the validity of this experiment.

\begin{itemize}
    \item \textbf{Internal validity.} One ongoing challenge with manually created benchmarks, such as \mybench, is the potential inclusion of extraneous noise, including typographical errors and unidentified bugs within the requirements, object-oriented models, and code. This can mislead LLMs and constitutes a threat to internal validity.
    To mitigate this issue, at least two authors jointly reviewed all artifacts contained in the benchmark to minimize the occurrence of noise.

    \item \textbf{External validity.} Given constraints in time and budget, we were unable to evaluate every state-of-the-art LLM, limiting the generalizability of our findings. To address this, we chose 7 LLMs from 3 distinct families, covering various parameter sizes, inference and non-inference models, as well as both open-source and closed-source models, to enhance representativeness.

    \item \textbf{Construct validity.} The experiment utilizes metrics such as \passk{} and \compk{}, consistent with existing studies. Although metrics like mutation scores for test case effectiveness exist, we couldn't explore all due to time constraints, but this will be tackled in future work.

    % \item \textbf{Conclusion validity.} 
\end{itemize}

%% file: sections/relatedwork.tex
\section{Related Work}\label{sec:relatedwork}
 
Most of the early benchmarks for LLM-based software development focused on function-level code generation \cite{HumanEval-passk,humanEvalX,MBPP,mathqa,PandasEval-NumpyEval,TorchDataEval} and test case generation \cite{contest,testeval,testgeneval,swtbench,clover}. Subsequent benchmarks started to focus on more programming languages \cite{MulyiPL-MBPP,crosscodeevaldiversemultilingualbenchmark,classEval}, higher problem complexity (such as project-level generation) \cite{repocoder,DevEval-BowenLi-etal-2025-prompting,JavaBench10.1145/3691620.3695470}, and more development stages \cite{DevEval-BowenLi-etal-2025-prompting}. \mybench mainly focuses on evaluating the ability of LLMs in understanding, performing, and obeying designs in different stages of software development, aiming to promote research and application of LLMs in complex software development scenarios.

% JavaBench: 
Regarding comparable benchmarks, JavaBench \cite{JavaBench10.1145/3691620.3695470} is a project-level Java benchmark that exercises OOP features with 4 projects.
It focuses on code generation based on class skeletons with detailed annotations.
The OOP benchmark \cite{oopEval-wang-etal-2024-oop} is a Python benchmark emphasizing object-oriented programming. Each problem presented offers a comprehensive description, detailing both functional requirements and the needed classes/methods---akin to a detailed design.
ProjectEval \cite{projecteval} contains 20 project-level code generation tasks. It is unique for providing 3 different levels of descriptions for each task, including a short NL description, an NL checklist for development, and a class skeleton.
\mybench stands out from these benchmarks in two key ways. Firstly, it is design-aware, allowing the evaluation of LLMs' software design comprehension through diverse input combinations. Secondly, it offers versatility by catering to code generation, design model generation, and test case generation tasks.

DevBench \cite{DevEval-BowenLi-etal-2025-prompting}, which consists of 22 project-level problems, covers the entire development workflow. However, all the problems in DevBench come from open-source projects. Among these, there are 13 problems that can be trivially found in GitHub by using the project names. In other words, LLMs can simply \textit{search} their databases, rather than reasoning, to return an almost correct answer, i.e., the data contamination issue.

%% file: sections/conclusion.tex
\section{Conclusion and Future Directions}\label{sec:conclusion}
This paper introduces \mybench for evaluating the capabilities and limitations of LLMs in executing design-related SE tasks, including object-oriented modeling, and the generation of code and test cases with/without design specifications.
\mybench includes 30 Java projects independently developed without relying on open-source projects or existing benchmarks.
We demonstrate the usage of \mybench by evaluating seven state-of-the-art LLMs against \mybench.
Our experimental findings indicate that, at present, LLMs lack the capability to manage the intricate nature of software design. 
The replication package includes all data of \mybench, scripts, raw results, and some additional analysis documents. 

In view of the current gaps revealed by \mybench, we further propose theree perspectives for future research:

\begin{itemize}
    \item \textbf{Investigating LLM-compatible formats for software design}. Current LLMs face challenges in comprehending and interpreting software design specifications, particularly the intricacies of software design models (encoded in PlantUML format in this paper). While it is possible to relay software design to LLMs using NL, this approach often involves unstructured and vague expressions that may introduce unwanted variability. Consequently, it is beneficial to pursue the development of a software design format optimized for LLM compatibility.

    \item \textbf{Integrating rule-based and AI-base code generation}. In model-driven and low-code development, various rule-based code generation techniques are available to transform software design models into code (skeletons). Since LLMs have difficulty understanding abstract software models, a practical strategy is to combine rule-based and AI-based methods: using rule-based approaches first to convert design models into initial code, then employing LLMs to finish the incomplete functions. It is required to investigate the effective integration of these methods and the use of design models to guide LLMs in enhancing the initial code.

    \item \textbf{Reasoning for software designs}. To support complex software development, it is critical to explore new LLM-driven software design methodologies. Such methodologies would not only guide LLMs in modularizing systems and planning development tasks but also identify subtask dependencies and enrich prompts with contextual information. Our experiments demonstrate that reasoning models exhibit strong design capabilities; however, despite these strengths, they still struggle with recognizing operations and inter-class relationships, a critical gap in object-oriented design. Therefore, advancing LLMs' software design proficiency through fine-tuning or specialized training remains a key research priority for future AI-driven SE.

\end{itemize}